\documentclass[useAMS,usenatbib,pdftex,letterpaper]{mn2e}
\usepackage{graphicx,epsfig} 
\usepackage{color}

\title[The wind of the pre-FUor V1331 Cyg]
    {Facing the wind of the pre-FUor V1331 Cyg 
    \thanks{Based on observations collected at the Keck Observatory, Hawaii, USA; the Centro Astron\'omico Hispano Alem\'an (CAHA) at Calar Alto, operated jointly by the Max-Planck Institut f\"ur Astronomie and the Instituto de Astrof\'{\i}sica de Andaluc\'{\i}a (CSIC); the WHT telescope on La Palma, Spain; and the Crimean Astrophysical Observatory, Ukraine.}}
		
\author[Petrov, Kurosawa, Romanova et al.]
  {P.P.~Petrov $^{1}$\thanks {E-mail:petrov@crao.crimea.ua; petrogen@rambler.ru},
  R.~Kurosawa$^{2}$, M.M.~Romanova$^{3}$, J.F.~Gameiro$^{4}$, M.~Fernandez$^{5}$,
	\newauthor E.V.~Babina$^{1}$ and  S.A.~Artemenko$^{1}$\\ 
       $^{1}$Crimean Astrophysical Observatory, Taras Shevchenko National University
              of Kiev, 98409 Nauchny, Crimea,  Ukraine\\
      $^{2}$Max-Planck-Institut f\"{u}r Radioastronomie, Auf dem
              H\"{u}gel 69, 53121 Bonn, Germany\\
      $^{3}$Cornell University, 310 Space Sciences Building, Ithaca
              NY, 14853 USA\\
      $^{4}$Centro de Astrof\'{\i}sica e Faculdade de Ci\^encias da Universidade do Porto,
              Rua das Estrelas, 4150-762 Porto, Portugal\\
      $^{5}$Instituto de Astrof\'{\i}sica de Andaluc\'{\i}a, CSIC, Glorieta de la
              Astronom\'{\i}a, 3, E-18008 Granada, Spain}

\date{ June 19, 2014}

\pagerange{\pageref{firstpage}--\pageref{lastpage}} \pubyear{2014}

\begin{document}
\label{firstpage}
\maketitle

%  \label{firstpage}
  \begin{abstract}

   The mass outflows in T Tauri stars (TTS) are thought to be an effective mechanism
to remove angular momentum during the pre-main-sequence contraction of a low-mass star.
The most powerful winds are observed at the FUor stage of stellar evolution. 
V1331 Cyg has been considered as a TTS at the pre-FUor stage.
 We analyse high-resolution spectra of V1331 Cyg collected in 1998--2007 and
20-d series of spectra taken in 2012. 
 For the first time the photospheric spectrum of the star is detected and 
stellar parameters are derived: spectral type G7--K0 IV, mass 2.8 M$_\odot$, 
radius 5 R$_\odot$, $v\,\sin i$ $<$ 6 km\,s$^{-1}$.
%The radial velocity RV = -15.0 $\pm$ 0.3 km\,s$^{-1}$.
The photospheric spectrum is highly veiled, but the amount of veiling is not
the same in different spectral lines, being lower in weak transitions and much
higher in strong transitions. The Fe\,{\sc ii} 5018, Mg\,{\sc i} 5183, K\,{\sc i} 7699 and some other lines of metals are accompanied by a `shell' absorption at radial velocity of about -240 km\,s$^{-1}$.
We show that these absorptions form in the post-shock gas in the jet, i.e. the star 
is seen though its jet.
The P Cyg profiles of H$\alpha$ and H$\beta$ indicate the terminal wind velocity 
of about 500 km\,s$^{-1}$, which vary on time-scales from several days to years.
A model of the stellar wind is developed to interpret the observations. The model is based on calculation of hydrogen spectral lines using the radiative transfer code 
{\sc torus}. 
The observed H$\alpha$ and H$\beta$ line profiles and their variability can be well reproduced with a stellar wind model, where the mass-loss rate and collimation 
(opening angle) of the wind are variable. The changes of the opening angle may be induced by small variability in magetization of the inner disc wind.
The mass-loss rate is found to vary within 
(6$-$11)$\times$10$^{-8}$ M$_\odot\,\mathrm{yr}^{-1}$, 
with the accretion rate of 2.0$\times$10$^{-6}$ M$_\odot\,\mathrm{yr}^{-1}$.

\end{abstract}
 
%________________________________________________________________

\section{Introduction}
\label{sec:intro}
T Tauri stars (TTS) are pre-main-sequence (PMS) objects of low masses ($\leq$ 2 M$_\odot$)
at ages of $\sim$1--10 Myr. The classical TTS (cTTS) still possess their
accretion discs. The processes of mass accretion on to cTTS are responsible
for the observed irregular light variability  and the intensive emission line spectra
of the stars. The accretion is also thought to be the driving force of the observed
mass outflows -- winds and jets of cTTS.
For review of the observational characteristics of cTTS and their models, see
e.g.~\citet{Bouvier07} and \citet{Guenther13}.

One of the intriguing problems of cTTS is the evolution of their angular momentum.
In spite of high angular momentum of the accreting matter, most of cTTS rotate
at less than 0.1 of their critical velocity, with periods of several days.
There must be some mechanism to remove the excess of angular momentum from
the star--disc system during the first million years of their evolution. 
The winds and jets formed in magnetohydrodynamic (MHD) processes are the
probable agents through which cTTS lose their angular momentum.
The large-scale open magnetic field connects the rotating star--disc system
with the circumstellar medium, and the magnetized outflowing gas removes
the mass and angular momentum from the system \citep{Matt05}.

Different configurations of the wind formation are possible, including a
stellar wind similar to the solar one \citep{Cranmer09}, an X-wind, in which
the outflow starts from the inner region of the disc \citep*{Shu04},
and a disc wind launched from the extended disc area \citep{Pudritz07}.
A conical wind model, launched from the inner disc and accelerated by magnetic pressure,
was proposed by \citet{Romanova09}.

Observations may provide clues to the origin of outflows and constrain the models.
Comparison of the observed diagnostic line profiles with those predicted
by the models is a usual tool in the study of the accretion and outflow processes
in cTTS (e.g.~\citealt{Edwards06}). Major recent efforts are from
\citet*{Kurosawa11} and \citet{Kwan11},
who studied the effect of the winds on the formation of hydrogen and helium lines
in optical and near-infrared, using their radiative transfer model.

The mass-loss rates in cTTS are typically within 
10$^{-9}$~--~10$^{-7}$~M$_\odot\,\mathrm{yr}^{-1}$.
Extreme case of mass-loss can be seen in the FU Ori stars (FUors). This stage of the
PMS evolution may be considered as a dramatic episode of intensified
transfer of angular momentum. It is widely agreed that the FUor phenomenon
is an event of greatly enhanced accretion, and the intensive outflow is a consequence
of this \citep{Hartmann96, Audard14}. 
However, in one of the classical FUor, V1057 Cyg, which was a cTTS before its brightening in 1971, an extremely powerful wind already was present in 1958, about 12 years {\it before} the outburst 
\citep{Herbig09}.
This suggests that the FUor progenitors are cTTS with enhanced mass-loss.

Among the cTTS there is a small group of stars which possess unusually strong winds,
similar to those in FUors, and may possibly be progenitors of FUors. Apart from the
typical P Cyg profiles, an obvious indicator of their dense, high-velocity winds is
the abnormal ratio of the H and K Ca\,{\sc ii} emission lines: while the K line
of Ca\,{\sc ii} (3933\,\AA) is prominent in emission, the H line (3968\,\AA)
is absent because it is suppressed by the P Cyg absorption component of H$\epsilon$
(3979\,\AA)~\citep*{Herbig03}.
In other words, the wind is optically thick even in higher Balmer lines.
This effect was also seen in the spectrum of V1057 Cyg before it went to FUor stage.

Three such extreme-wind cTTS, that are also relatively bright (V$<13^m$), can be
identified in the northern sky: V1331 Cyg, AS 353 A and LkHa 321.
The study of these stars may shed light on the
nature of the transition between the cTTS and FUor phases of the PMS evolution.

In this paper we present results of our research of V1331 Cyg.
The star was earlier considered as a candidate in pre-FUors 
\citep*{Welin76, Herbig89, MacMuldroch93}.
Besides the strong emission line spectrum
and strong wind features, the star is surrounded by a ring-like reflection
nebula of about 30 arcsec in diameter \citep{Kuhi64, Mundt98}.
Such nebulae are present in classical FUors, indicating the past events of extensive
mass-loss \citep{Goodrich87}. From the images of V1331 Cyg obtained by the \textit{Hubble Space Telescope},
\citet*{Quanz07} revealed yet another ring-like nebula closer to the star.
They concluded that the star is seen pole-on, along the axis of a conical outflow.
Radio emission of CO molecule in mm wavelengths showed more complicated
structure: a massive ($\approx$0.5 M$_\odot$) disc around the star, bipolar flows and
an expanding ring of about 10$^4$~au \citep{MacMuldroch93}.
It was concluded that the previous FUor event of V1331 Cyg was about 4000 years ago.

Interestingly, the photospheric spectrum of the star has not been detected so far,
and the spectral type and luminosity have been estimated from its spectral energy distribution,
interstellar extinction and distance. Earlier estimations were B0.5 \citep{Cohen79},
A8--F0 \citep{Chavarria81}, F0 \citep{Mundt81},
G0 \citep{Kolotilov83}.
Later investigation by \citet{Hamann92} gave spectral type G5 and stellar
luminosity L$_*$ = 21 L$_\odot$, with distance d = 700 pc and extinction Av = 1.4$^m$.
The low dispersion IUE spectrogram at $\lambda$\,2200--3200\,\AA\, is dominated
by the Mg\,{\sc ii} resonance doublet emission \citep{Mundt81}. No Balmer jump in
emission is in the blue part of the spectrum \citep*{Valenti93}.
The visible region shows low excitation emission line spectrum of metals
and the P Cyg features of Balmer lines of hydrogen.
With near-IR interferometry, the dusty disc inner radius (at the distance of
dust sublimation) was measured as 0.31 au \citep{Eisner07}.
V1331 Cyg is photometrically variable within $V$=11.8--12.4 
\citep{Kolotilov83, Fernandez96, Shevchenko03}. No rotational period was found from the available
photometrical data.

The aim of our research is to find an adequate model of wind of V1331 Cyg,
which can describe the observed Balmer line profiles and their variability.
We use high-resolution, high quality echelle spectra of V1331 Cyg, obtained in
1998, 2004 and 2007, and a series of spectra obtained in 2012 August.

%%%%%%%%%%%%%%%%%%%%%%%%%%%%%%%%%%%%%%%%%%%%%%%%%%%%%%%%%%%%%%%
\section{Observations}
\label{sec:obs}

One spectrum of V1331 Cyg was obtained at the 4.2m \textit{William
  Herschel Telescope} of the Isaac Newton Group, using the Utrech
Echelle Spectrograph (UES), equipped with an echelle grating of 31
lines per mm and installed on the Nasmyth focus. The instrument
yielded 67 orders spanning a wavelength range of $\approx$ 4650--10100
\AA. A SITe2 chip 2048$\times$2048 pixel CCD detector with  24$\mu$m
pixel was used. The spectral resolution  R $\approx$ 50000  and the
signal-to-noise ratio (S/N) is about 150 at 6500 \AA. 
%Standard reduction was performed using IRAF package.

Two spectra of V1331 Cyg were obtained by George Herbig with the HIRES
echelle spectrograph at Keck-1\footnote{The W. M. Keck Observatory is
  operated as a scientific partnership among 
the California Institute of Technology, the University of California, and the
National Aeronautics and Space Administration. The Observatory was made
possible by the generous financial support of the W. M. Keck Foundation.}
%%%%%%
on 2004 July 24 and 2007 November 23. In 2004, the CCD detector covered wavelength
range of 4350--6750 \,\AA. In 2007, a mosaic of three CCDs was used to cover the range of
4750--8690 \,\AA. The data have the spectral  resolution R $\approx$ 48000, and the S/N=150--250.  
%The spectra were reduced in a standard way with IRAF package.

In 2012, we carried out spectroscopic monitoring of V1331 Cyg during about 20~d
with the aim to detect possible rotational modulation in emission line profile,
which could help us to estimate the period of stellar rotation and characteristic
time of the wind profiles variability.
The observations were carried out in two observing sites:
Calar Alto Observatory (Spain) and Crimean Astrophysical Observatory (Ukraine).
The log of observations is given in Table~1.

Eleven spectra were obtained from July 28 to August 22, with the Calar Alto 
Fibre-fed Echelle (CAFE) attached to the
 Cassegrain focus of the 2.2m Telescope of the Calar Alto Observatory
 (CAHA). CAFE is a single-fibre, high-resolution (R $\approx$ 60000)
 spectrograph, covering the wavelength range of 3650--9800~\AA\/ with 84
 orders \citep{Aceituno13}. The detector is an iKon-L camera
 with 2048x2048 pixels of 13.5~$\mu$m. 
%The aperture is limited by the entrance diaphragm, which has an aperture of 200~m$\mu$ (ie., 2.4 arcsec  projected in the sky). 
 In this set of spectra the S/N varies from
 night to night and its value for the continuum near the H$\alpha$ line
 is between 20 and 35. 
 %Standard data reduction was performed using the IRAF package.

Observations at Calar Alto were immediately followed by the run of
the Crimean observations with the coud\'{e} spectrograph of the 2.6-m Shajn reflector.
Six consequent 30-min exposures were taken each night.
The S/N in the Crimean spectra was lower than that in the Calar Alto spectra,
and only H$\alpha$ region (6530--6600\,\AA) was covered.
In order to increase the S/N, the Crimean spectra were smoothed by Gaussian
filter with full width at half-maximum (FWHM) = 0.3\,\AA\,(13.7 km\,s$^{-1}$).

All the spectra were reduced in a standard way using the IRAF routines 
and nomalized to continuum level. 
The spectral monitoring in 2012 was supported by UBVR photometry at the
Crimean Astrophysical Observatory.

\begin{table}
\caption{Log of observations}
\label{Tab1}
\begin{tabular}{|l|c|l|c|}
\hline
\bf Site &  Year & Date & Mid exposure HJD 245...\\ 
\hline
La Palma  & 1998 &Nov 7 & 1125.331\\
Mauna Kea & 2004 &Jul 24 & 3210.553 \\
\smallskip
Mauna Kea & 2007 &Nov 23 & 4427.726  \\
Calar Alto & 2012 &Jul 28&  6136.655 \\
           &      &Aug 13& 6152.518 \\
           &      &Aug 14& 6153.548 \\
           &      &Aug 15& 6154.640 \\
           &      &Aug 16& 6155.525 \\
           &      &Aug 17& 6156.677 \\
           &      &Aug 18& 6157.675 \\
           &      &Aug 19& 6158.653 \\
           &      &Aug 20& 6159.681 \\
           &      &Aug 21& 6160.676 \\
\smallskip					
           &      &Aug 22& 6161.680 \\

Crimea  &  2012 &Aug 21&  6161.277 \\
        &       &Aug 22&  6162.407 \\
        &       &Aug 23&  6163.268 \\
        &       &Aug 24&  6164.410 \\
        &       &Aug 25&  6165.249 \\
        &       &Aug 26&  6166.413 \\
        &       &Aug 27&  6167.237 \\
        &       &Aug 30&  6170.445 \\
        &       &Aug 31&  6171.253 \\
\hline
\end{tabular}
\end{table}

%%%%%%%%%%%%%%%%%%%%%%%%%%%%%%%%%%%%%%%%%%%%%%%%%%%%%%%%%%%%%%%%%%%%%%
\section{Results}
\label{sec:results}

Two fragments of spectrum of V1331 Cyg are shown in Fig~\ref{fig1}. In this and the following figures, the wavelength scale is astrocentric.
As in many other cTTS (e.g.~\citealt{Hamann92}), 
the spectrum of V1331 Cyg consists of several components:\\
%(e.g. ~\citet{Petrov03}, ~\citet{}):\\
1) narrow veiled photospheric absorptions of a late-type star;\\
2) narrow emission lines of metals -- neutrals and ions;\\
3) lines of Balmer and Pashen series, with P Cyg profiles;\\
4) narrow deep absorptions, blue-shifted by 150--250 km\,s$^{-1}$, in lines of
     Fe\,{\sc ii}, Mg\,{\sc i}, Na\,{\sc i} and others;\\
5) forbidden emission lines of [O\,{\sc i}] and [S\,{\sc ii}].\\
In the following we consider in details each of these components.

\begin{figure}
\centerline{\resizebox{8.5cm}{!}{\includegraphics{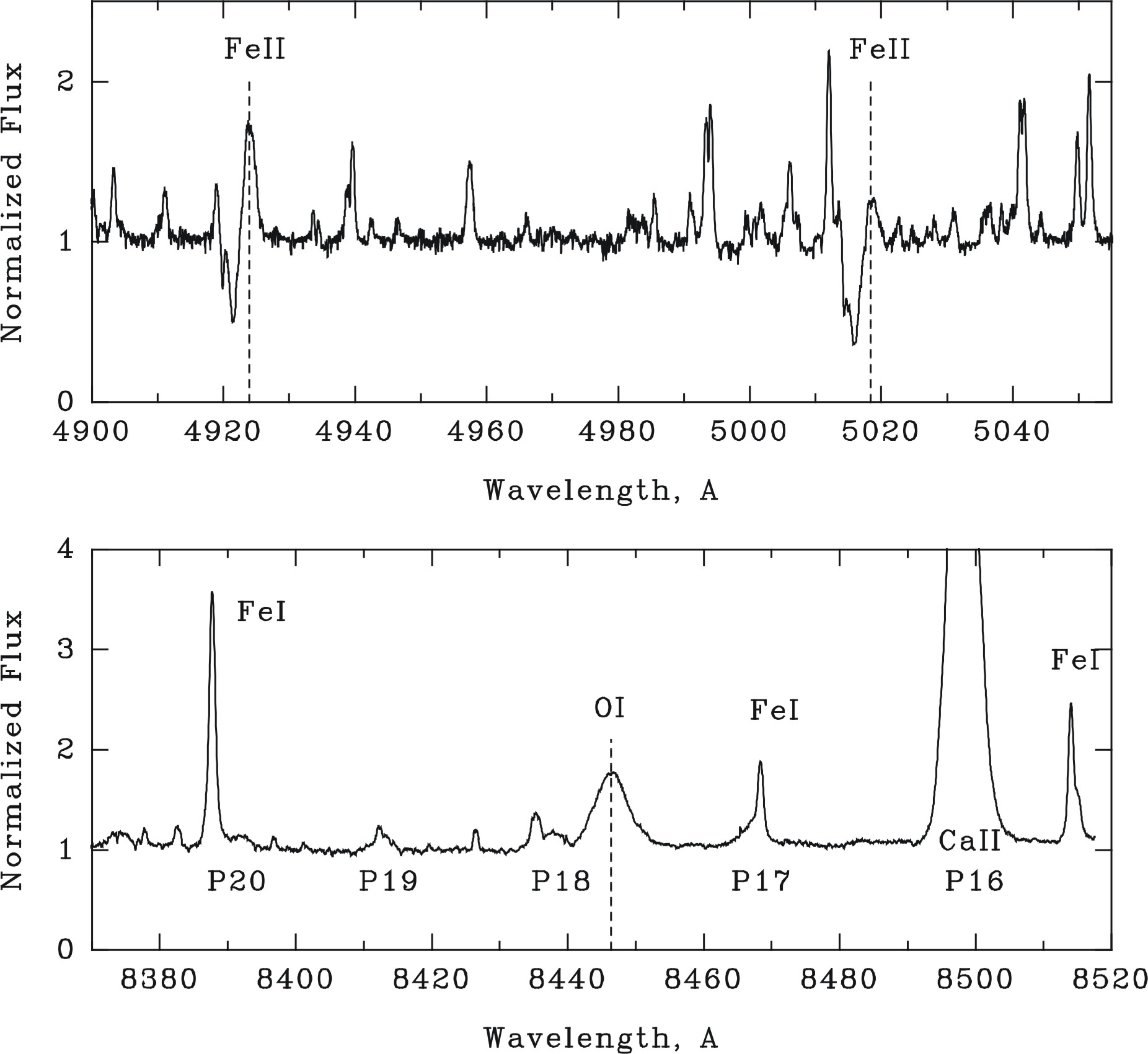}}}
\caption{Two fragments of spectrum of V1331 Cyg in 2007.
       Wavelength scale is astrocentric. The vertical dashed lines mark
       laboratory wavelengths of the Fe\,{\sc ii} and O\,{\sc i} lines.}
\label{fig1}
\end{figure}

\subsection{Photospheric spectrum and stellar parameters}
\label{sec:photospheric}
\begin{figure}
\centerline{\resizebox{8.5cm}{!}{\includegraphics{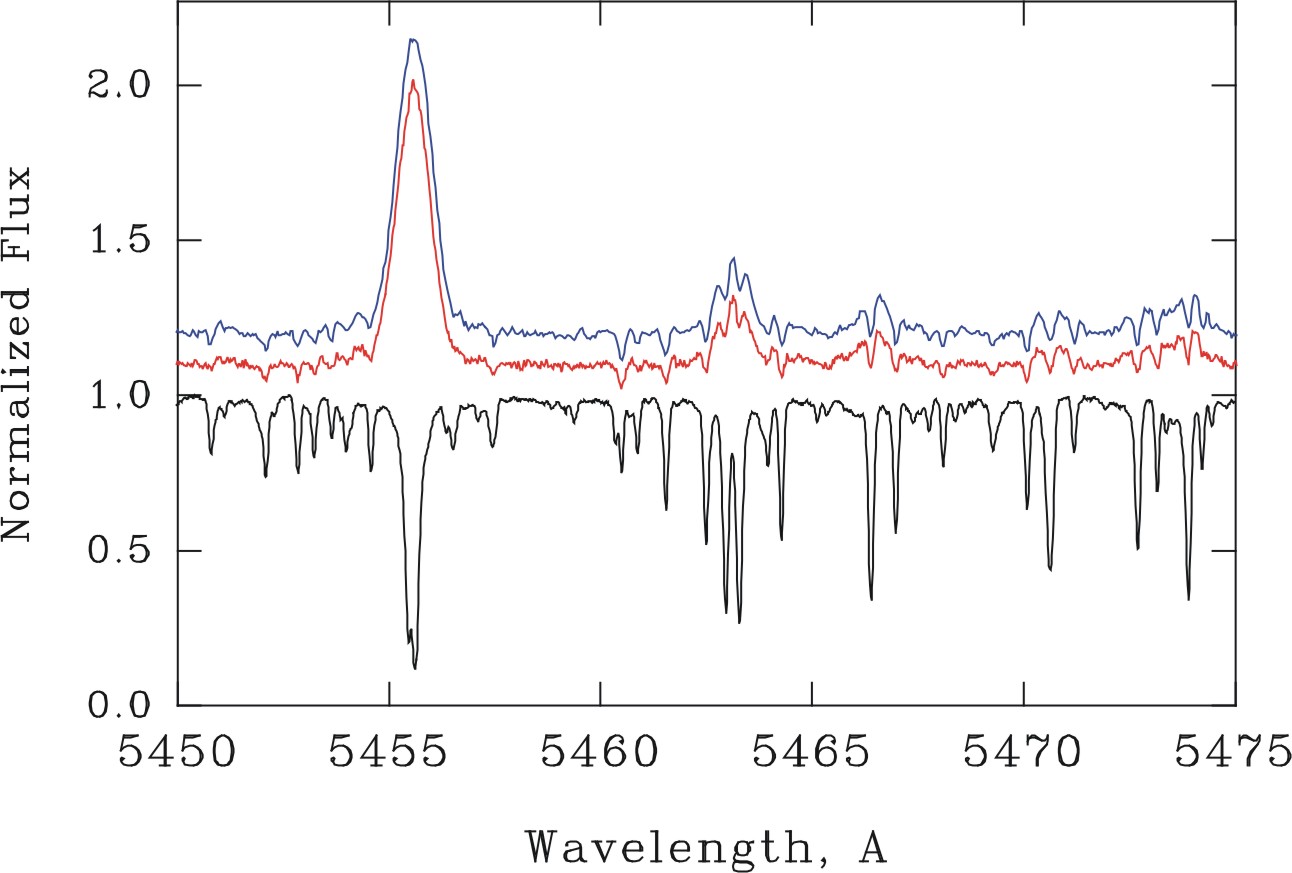}}}
\caption{An example of the weak photospheric lines in V1331 Cyg, 2004 and 2007
(two upper curves). A template G7 IV spectrum is shown for comparison (lower curve).}
\label{phot_sp}
\end{figure}

The photospheric spectrum of the star is best visible in the Keck spectra (Fig.~\ref{phot_sp}), although the lines are very weak and narrow.
In spite of the high resolution (6 km\,s$^{-1}$), the photospheric lines are not resolved:
their width is the same as that of the weak telluric water lines. Thus, we may set
only the upper limit for the projected rotational velocity of the star:
$v\,\sin i$ $<$ 6 km\,s$^{-1}$.
This is in agreement with the earlier conclusion that the star is seen pole-on.
The radial velocity (RV) of the star RV = -15.0 $\pm$ 0.3 km\,s$^{-1}$, with no difference between
the spectra of 2004 and 2007.

The spectral classification of the highly veiled spectrum of cTTS is not
a trivial task. The line ratios may be distorted by the chromospheric emission
filling in the stronger lines, therefore the temperature and gravity criteria
must be found among the weakest lines.
We compared the photospheric spectrum of V1331 Cyg with a number of spectra
downloaded from the VLT/UVES library
 \footnote{http://www.eso.org/sci/observing/tools/uvespop/},
within spectral types G5--K2 and luminosities II--IV.
Also available was the spectrum of $\beta$\,Aqr (G0 Ib--II), taken with the HIRES
spectrograph at Keck-1.
In addition, we used a grid of synthetic spectra in selected wavelength
windows in order to find luminosity criteria.
 The spectra were calculated using the code by Berdyugina (1991) and
Kurucz models. Atomic line data were retrieved from the
VALD data base (Kupka et al. 2000).
More detailed description of the spectral type and luminosity determination
is given in a separate paper~\citep{Petrov14}.
As a result, the spectral type was found to be within G7--K0 IV,
which corresponds to $T_{\rm eff}$ = 5000--5250 K,  log $g$ $\approx$ 3.5.
This value of gravity indicates that the observed photospheric spectrum
 of V1331 Cyg
is formed not in the disc atmosphere, as it could be in case of a FUor, but in
the atmosphere of the star. 
With this temperature, assuming stellar luminosity
21~L$_\odot$ (Hamann \& Persson, 1992) the mass and radius of the star
were derived using the grid of models by \citet*{Siess00}:
$M_*$~$\approx$~2.8~M$_\odot$, $R_*$~$\approx$~5~R$_\odot$.

\subsection{The peculiar veiling effect}
\label{subsec:veiling}

%Special attention should be given to the veiling effect in V1331 Cyg.
The veiling of the photospheric spectrum in cTTS is usually attributed
to the presence of an additional (non-photospheric) continuum, radiated by
a hotspot on stellar surface. 
Then, the veiling factor (VF)~=~EW(std)/EW(tts)~$-$~1,
where EW(tts) is equivalent width of a line in spectrum of TTS, and EW(std)
is that in a standard star of the same spectral type. The VF is typically
wavelength dependent, rising towards the blue part of the spectrum.
In this interpretation, all the photospheric lines within a narrow spectral range
must be reduced in EW by the same factor. However, detailed analysis of
highly veiled cTTS spectra revealed that stronger lines are more affected by
veiling than weaker lines, even those close in wavelength 
\citep{Gahm08, Gahm13, Petrov11}. It was interpreted as an effect of
chromospheric emission filling in stronger lines,
and was reproduced in a model of atmosphere heated by 
accretion \citep{Dodin12}. 

\begin{figure}
\centerline{\resizebox{7cm}{!}{\includegraphics{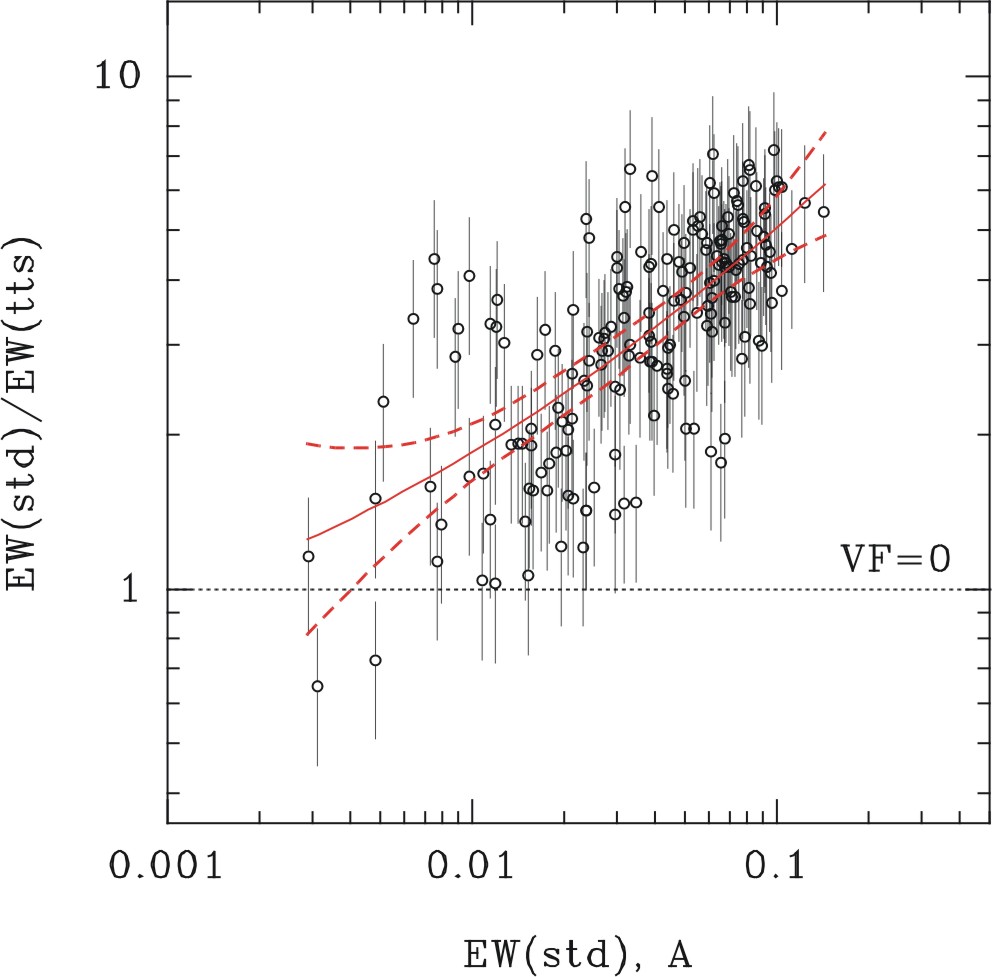}}}
\caption{Ratio of equivalent widths in spectra of template (G7 IV) and V1331 Cyg
      is rising towards stronger lines. The line of regression and the
      99 per cent confidence interval are indicated. The
      horizontal dotted line marks zero level of veiling.} 
\label{veil}
\end{figure}

This effect of the `chromospheric veiling' is well expressed in V1331 Cyg.
We measured equivalent widths of about 200 photospheric lines in
V1331 Cyg and in the template stars. Fig.~\ref{veil} shows the ratio of EWs
as a function of EW in the template star. Stronger lines, with EW = 50--100 m\AA\,
in the template spectrum, are reduced by a factor of $\approx$4 in V1331 Cyg,
while the weaker lines are almost the same. The errors of EW measurements
in V1331 Cyg are large, because the measured lines are very weak, from a few m\AA\,
to about 20 m\AA. In the weakest lines (EW $<$ 5 m\AA), the error is 30--50 per cent,
for stronger lines it is about 20--30 per cent and caused mainly by the continuum
level uncertainty.

The dependence shown in Fig.~\ref{veil} still remains if another star, K1 IV, is used
as a template spectrum. There was no sense to use earlier G-type templates, because the
VF would became negative in weaker lines. Later K-type templates
are also not adequate, because already in K3 star numerous metal lines of low
ionization appear, which are certainly not present in V1331 Cyg.
Another complication is the presence of the emission lines of metals:
in stronger lines the photospheric absorption appears only as a dip on top of the
broader emission (see Fig.~\ref{phot_sp}). These lines were not included into analysis of EWs.

Hence, the VF, caused by a non-photospheric continuum in V1331 Cyg,
as derived from the weaker photospheric lines (EW $<$ 10 m\,\AA\, in the template star),
is not well defined, but certainly does not exceed VF = 1.
With this reservation, we do not find dependence of VF on either wavelength
(from 4500 to 8500\,\AA) or excitation potential (EP) (from 0 to 6 eV) of the transitions.
 
%%%%%%%%%%%%%%%%%%%%%%%%%%%%%%%%%%%%
\subsection{Emission line spectrum}

The emission line spectrum of V1331 Cyg is very rich in strong narrow
(FWHM = 40--60 km\,s$^{-1}$) lines of neutral and ionized metals, rested at stellar
RV. Intensities of the narrow emission lines are similar in
all the spectra of 1998--2012.
The narrow emission lines in spectra of cTTS are usually attributed to
chromospheric-like regions of post-shocked gas at the footpoints of accretion
columns (\citealt{Batalha96}; \citealt*{Beristain98}).
In the Keck spectra of V1331 Cyg, we measured EWs of 32 less blended emission lines
of Fe\,{\sc i} with EP of lower level from 0.9 to 4.5 eV,
and EWs of seven lines of Fe\,{\sc ii} with EP from 2.9 to 3.9 eV. The curve of growth
of these emission lines \citep{Herbig90} gives $T_{\rm exc}$ = 3800 $\pm$ 300 K and
log $N_{\rm e}$ = 8 $\pm$ 0.5.
This low electron density implies that the origin of the narrow emission lines
in V1331 Cyg may be different from those in cTTS. The constancy of the emission lines
also suggests their origin in a large volume of gas.

The broad emission lines are seen in the Balmer and Paschen series of hydrogen 
and the infrared Ca\,{\sc ii} triplet. Also broad lines, centred at 
stellar velocity, are those of high EP, e.g.~He\,{\sc i} 5876\,\AA\,
(EP = 23 eV, FWHM$\approx$130 km\,s$^{-1}$) and 
O\,{\sc i} 8446\,\AA\, (EP = 9.5 eV, FWHM$\approx$200 km\,s$^{-1}$).
The He\,{\sc i} 5876 emission is relatively weak, EW = 0.2\,\AA.
The He\,{\sc ii} 4686 line is absent.
In our spectra of V1331 Cyg, the observed ratio 
Ca\,{\sc ii} 8498 : O\,{\sc i} 8446 = 6.5 $\pm$ 1.0 
corresponds to $T_{\rm e}$ $\approx$ 8000 K and log $N_{\rm H}$ $\approx$ 10$^{12}$ cm$^{-3}$ 
~\citep{Kwan11}.
Most likely, the lines are formed in stellar magnetosphere.

%Broader emission lines are those of higher EP, e.g. He\,{\sc i} 5876\,\AA (EP = 23 eV,
%FWHM$\approx$130 km\,s$^{-1}$), O\,{\sc i} 8446\,\AA\, (EP = 9.5 eV, FWHM$\approx$200 km\,s$^{-1}$).
%Also broad are the lines of Balmer and Pashen series of H\,{\sc i} and the infrared
%triplet of Ca\,{\sc ii}. The He\,{\sc i} 5876\,\AA\, emission is relatively weak,
%EW $\approx$ 0.3\,\AA. The He\,{\sc ii} 4686\,\AA\, line is absent.

%Fig.~\ref{balmer} shows profiles of H$\alpha$ and H$\beta$ lines.

\begin{figure}
\centerline{\resizebox{8cm}{!}{\includegraphics{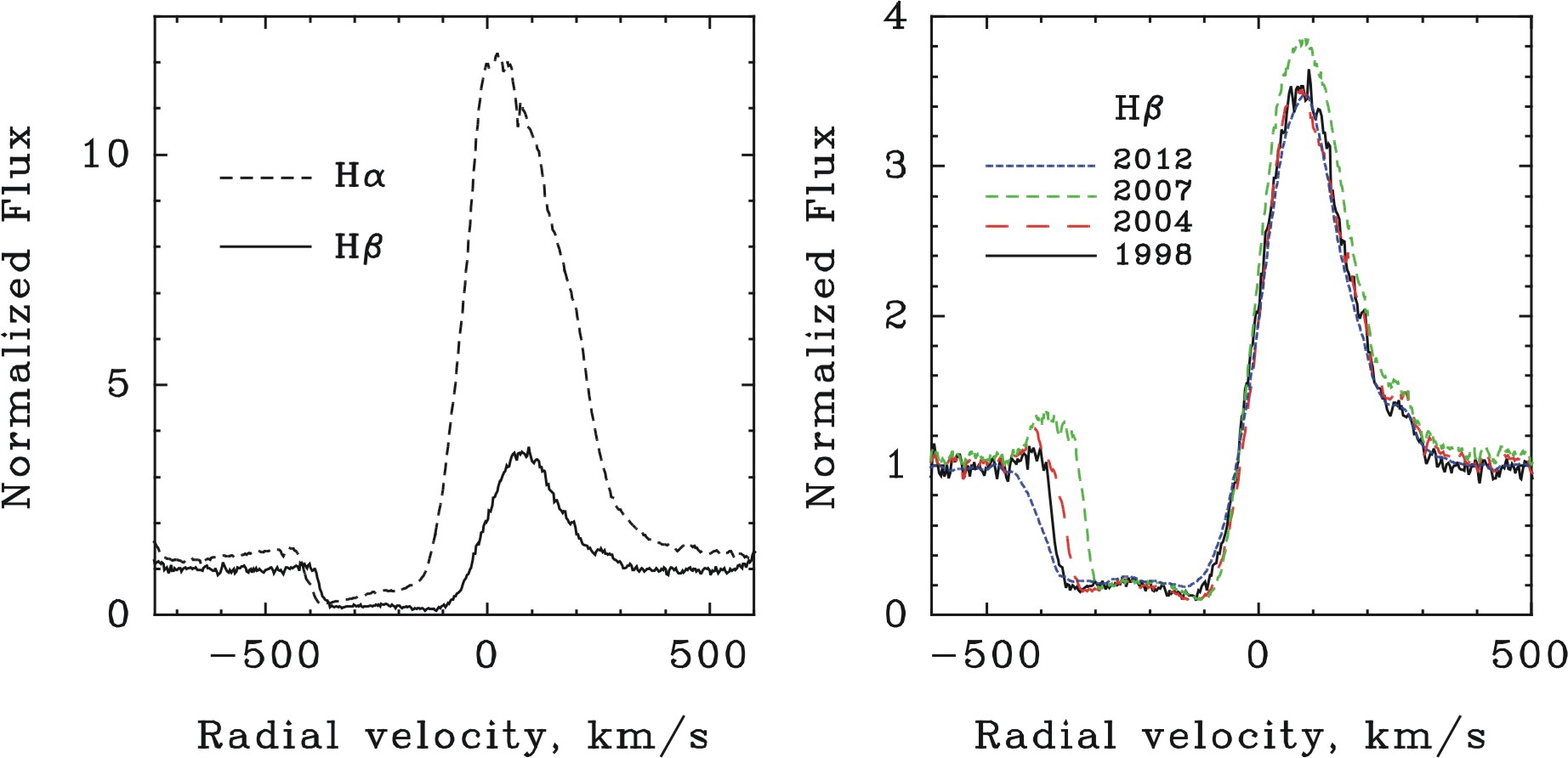}}}
\caption{Balmer line profiles. Left: H$\alpha$ and H$\beta$ lines in 2007.
        Right: H$\beta$ line in different years of observations.  }
\label{balmer}
\end{figure}

The forbidden emission lines of [O\,{\sc i}] 6300 and 6363\,\AA, and [S\,{\sc ii}]
6716 and 6730\,\AA\, have strong peaks at RV of $-240$~km\,s$^{-1}$. The peak position remained the same in all the years of observations, while the overall profile changed from year to year. 
These lines represent low density gas, $10^4$ to $10^6$ cm$^{-3}$ 
(e.g.~\citealt*{Hartigan95}). 
%The ratios of the forbidden lines correspond to an optically thin case.
%The [N\,{\sc ii}] lines are absent. 

%%%%%%%%%%%%%%%%%%%%%%%%%%%%%%%
\subsection{Wind features}
\label{sec:wind-features}

The Balmer lines of H\,{\sc i} show a classical P Cyg type profile, thus indicating
a powerful mass outflow (Fig.~\ref{balmer}). This characteristic is rare in cTTS but typical for
FUors. In our spectra of V1331 Cyg, the H$\alpha$ and H$\beta$ profiles are slightly
variable on a time-scale of days, while the spectra of different years show more
significant differences. The terminal RV of the outflow varies between 
-350 and -450 km\,s$^{-1}$. The strong P Cyg absorption is also present in the
resonance NaI D lines.

The characteristic pattern of variability is shown for H$\beta$ line on the right
panel of Fig.~\ref{balmer}, where profiles of 1998, 2004, 2007 and 2012 are overplotted.
The spectrum of 2012 is an average of 10 nights of Calar Alto observations.
The variable is mostly the terminal velocity of the wind, while the emission
peak remains about the same.

\begin{figure}
\centerline{\resizebox{8.5cm}{!}{\includegraphics{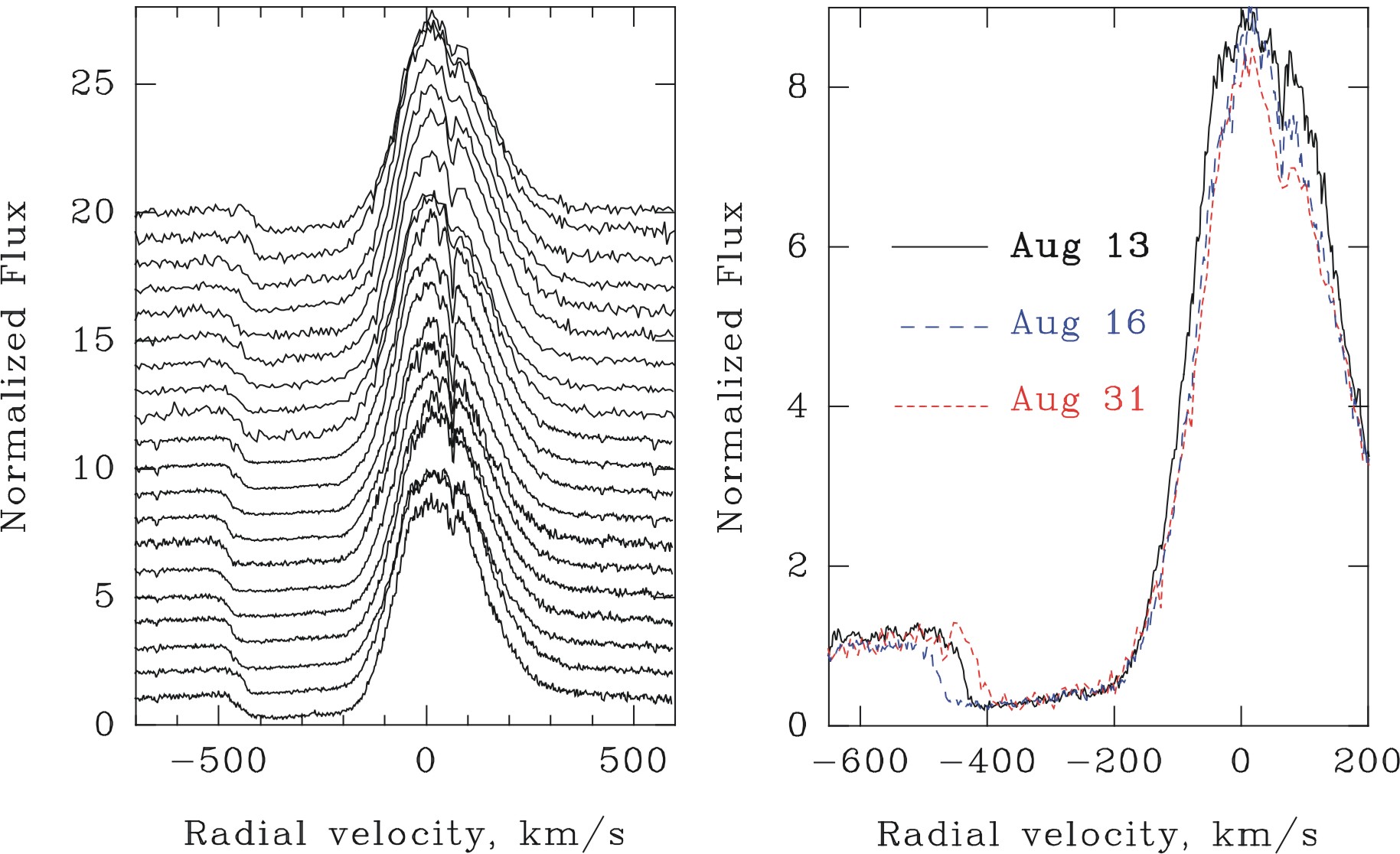}}}
\caption{Monitoring of H$\alpha$ line in 2012. Left panel: night-to-night
      series, starting from July 28 (lower spectrum).
      Right panel: three spectra of 2012 August, showing typical variability
      in the wind terminal velocity.}
\label{monitor}
\end{figure}

Fig.~\ref{monitor} shows a series of H$\alpha$ profiles in 2012 August, starting in Calar Alto
and continued in Crimea. The same kind of variability can be seen in this
20-d series: terminal velocity of the outflow changes by ~60 km\,s$^{-1}$ on
a time interval of a few days. This is shown with the three overplotted profiles
on the right panel of Fig 5.
No periodic variations, presumably related to stellar rotation, were found
in this 20-d monitoring.
Analysis of the Balmer line profiles is given in Section 4.

During the spectral monitoring in 2012 the brightness of the star varied slightly
within V = 11.85--12.08, B-V = 1.08--1.16. No correlation with any
spectral parameter was found.

%%%%%%%%%%%%%%%%%%%%%%%%%%%%%%%%%
\subsection{Accretion features}

\begin{figure}
\centerline{\resizebox{8.5cm}{!}{\includegraphics{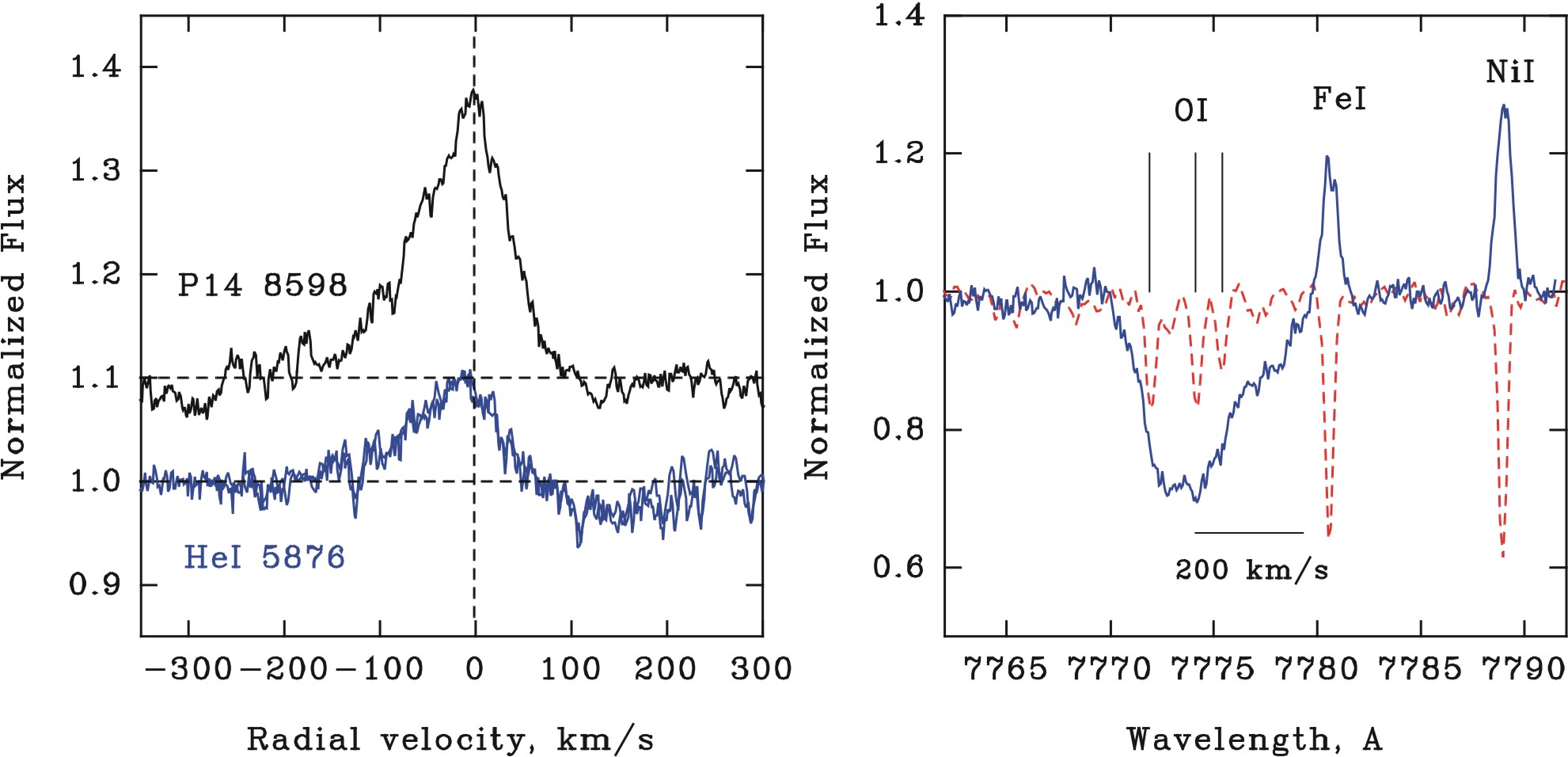}}}
\caption{Signatures of accretion in V1331 Cyg. Left panel: line profiles
       of He\,{\sc i} 5876\,\AA\ and  Pashen 14, shifted by 0.1 for clarity.
       Right panel: triplet of O\,{\sc i} (7772, 7774 and 7775 \,\AA) in $\beta$\,Aqr
       (dashed line) and V1331 Cyg (solid line). In V1331 Cyg, the red wing of the blend
       is extended to about 200 km\,s$^{-1}$.}
\label{IPC}
\end{figure}

The mass inflow is usually traced by the inversed P Cyg (IPC) profiles
of some diagnostic lines. In the optical region these are the higher Balmer lines,
the He\,{\sc i} lines and the triplet O\,{\sc i} 7773\,\AA. These indicators are strong in
actively accreting cTTS, but absent in spectra of FUors.
In V1331 Cyg, the IPC profiles are not well expressed, although noticeable in
the He\,{\sc i} 5876\,\AA, where the red wing of the emission is depressed by the
red-shifted absorption (Fig.~\ref{IPC}, left panel). A slight asymmetry can be noticed also
in the less blended Pashen~14 line.

Another indication is the broad absorption blend of the triplet O\,{\sc i} 7773\,\AA\,
(Fig.~\ref{IPC}, right panel).
This feature is present in the spectra of 1998 and 2012, but falls out of spectral order in
the Keck spectra of 2004 and 2007.
In both spectra of 1998 and 2012, the red wing of the absorption is extended to about
+200 km\,s$^{-1}$, which is in agreement with the He\,{\sc i} profile.
We may conclude that mass infall is going on in V1331 Cyg and the projected
infall velocity is about 200 km\,s$^{-1}$.

The apparent weakness of signatures of mass accretion in V1331 Cyg is probably
related to the pole-on orientation of the star. Besides of the mass accretion rate
and the viewing angle, the strength of the red-shifted absorption depends also 
on the size of magnetosphere and the gas temperature ~\citep{Muzerolle01}.
%For instance, noticeable red-shifted absorption in the He\,{\sc i} 5876\,\AA line wasobserved in other pole-on cTTS, DR Tau \citep{Petrov11} and RU Lup \citep{Gahm13}.

%%%%%%%%%%%%%%%%%%%%%%%%%%%%%%%%%%%%%%%
\subsection{`Shell' features}

\begin{figure}
\centerline{\resizebox{6.7cm}{!}{\includegraphics{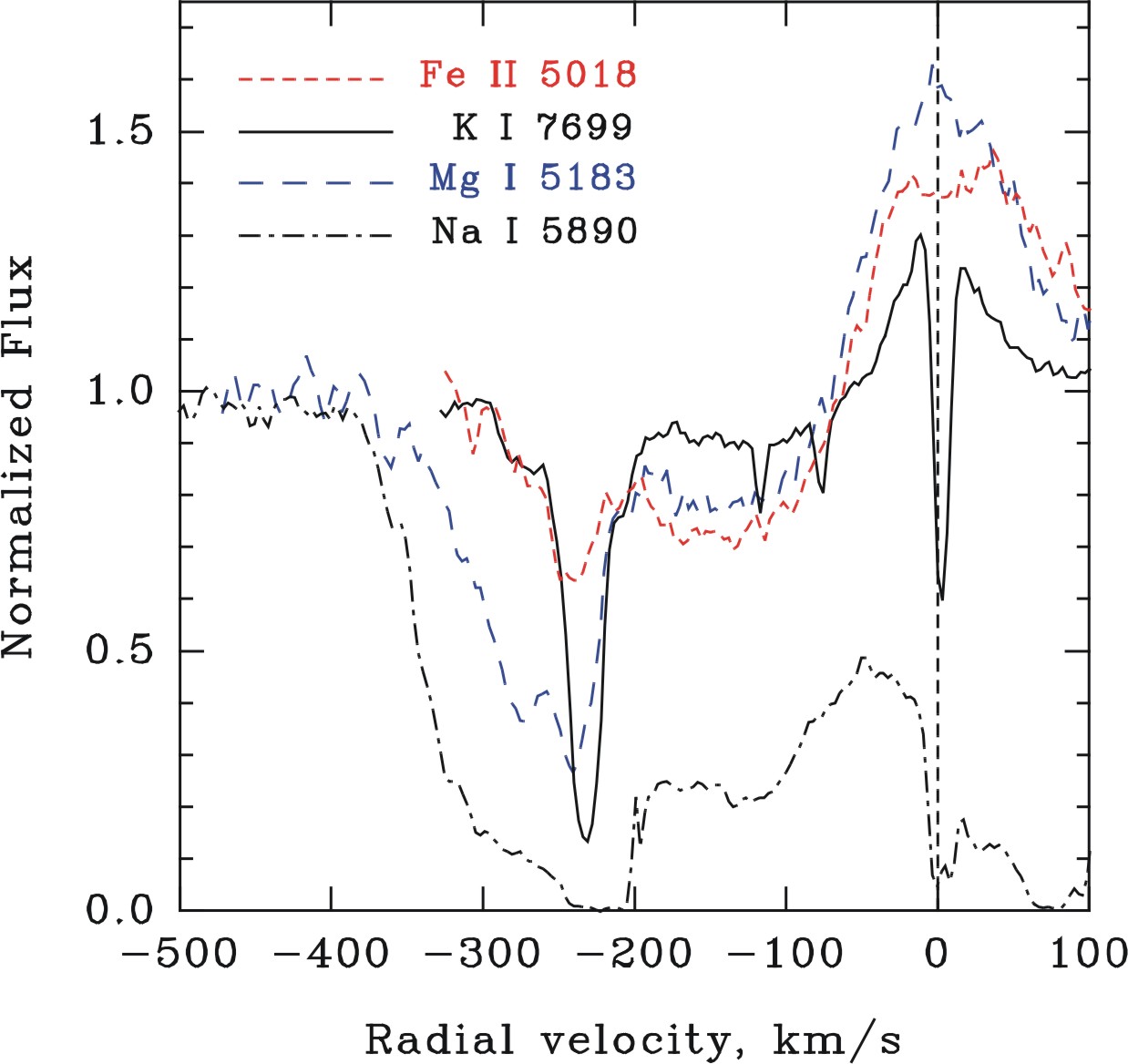}}}
\caption{The blueshifted  `shell' lines of Mg\,{\sc i}, Fe\,{\sc ii}, K\,{\sc i}
and Na\,{\sc i} with  maximum absorption at about -240 km\,s$^{-1}$. The spectrum of 1998.}
\label{shell}
\end{figure}

One peculiarity of V1331 Cyg spectrum is the presence of blue-shifted absorption
components of the emission lines of metals. These are so-called `shell' lines --
a signature of expanding gaseous shells. In our spectra of V1331 Cyg the `shell'
components are present in the following lines:
Fe\,{\sc ii} 4924 and 5118\,\AA, Mg\,{\sc i} 5183\,\AA, Li\,{\sc i} 6707\,\AA,
K\,{\sc i} 7699\,\AA, and Na\,{\sc i} D (see Figs~\ref{fig1} and \ref{shell}).
In the resonance line K\,{\sc i} 7699\,\AA, there is one distinct narrow `shell' feature
at -240 km\,s$^{-1}$. The same is present in the resonance Na\,{\sc i} doublet, although
saturated and blended with the wider P Cyg absorption.
In the Fe\,{\sc ii} and Mg\,{\sc i}, the `shell' profile is more complicated, although
the component at -240 km\,s$^{-1}$ is present there too. The component at -240 km\,s$^{-1}$
is stable over the years of our observations, while the overall profile varies
from year to year. No variability in the `shell' lines was found in the 20-d
period of our monitoring in 2012. We may conclude that the `shell' components
vary on a time-scale of a year. With the velocity of 240 km\,s$^{-1}$, this time of
variability corresponds to the distance scale of about 50 au.

\begin{figure}
\centerline{\resizebox{7.5cm}{!}{\includegraphics{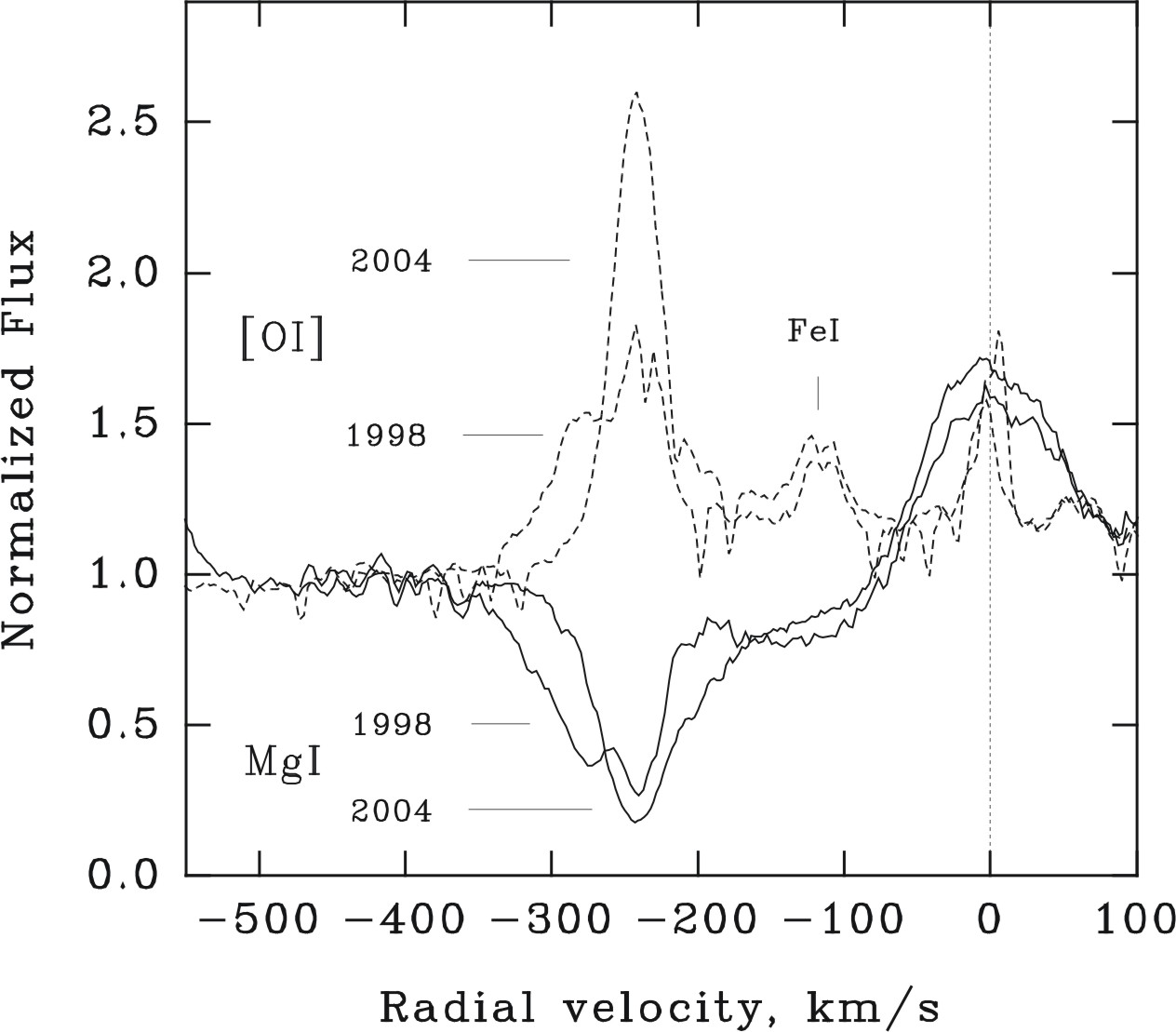}}}
\caption{Comparison of the forbidden [O\,{\sc i}] 6300\,\AA\,emission profile (dashed curves)
         and the blue-shifted `shell' absorption of Mg\,{\sc i} 5183\,\AA\, (solid line)
         in 1998 and 2004. The weaker emission peak at -120 km\,s$^{-1}$ belongs to
         Fe\,{\sc i} 6297.8 \,\AA.}
\label{shell_jet}
\end{figure}

There is a striking similarity in velocity profiles of the `shell' lines
and the forbidden lines (Fig.~\ref{shell_jet}). The Mg\,{\sc i} and [O\,{\sc i}]
profiles look like a  mirror reflection of each other.
We know that the blue-shifted component of the forbidden emission lines
is formed in jets of TTS (e.g.~\citealt{Hartigan95})
at large distance (tens of au) from the star, 
where ionized atoms recombine in the cooling region behind the shock.
The similarity of the `shell' profiles with those of the forbidden lines
strongly suggests that the `shell' absorptions also arise in the post-shocked
gas in the jet. With low inclination, we see the star through the jet, and
the line of sight (LOS) to the star intersects all the shocks in the jet.
The gas density is low, but the length scale is long enough to get
the column density of atoms necessary to form the `shell' absorption.
%For the [OI] emission the critical electron density is 10^6 cm^-3.

%%%%%%%%%%%%%%%%%%%%%%%%%%%%%%%%%%%
\section{Analysis: wind model}

%=========================================
\begin{table*}
  \caption{Adopted model parameters}
  \label{tab:model-par-common}
  \begin{center}
    \begin{tabular}{ccccccccccccc}
      \hline
      $R_{*}$ & $M_{*}$ & $T_{\mathrm{eff}}$ & $R_{\mathrm{mi}}$ &
      $R_{\mathrm{mo}}$ & $T_{\mathrm{m}}$ &
      $\dot{M}_{\mathrm{a}}$ & $T_{\mathrm{w}}$ & $v_{\infty}$ & $v_{0}$
      & $\beta$ & $R_{0}$\\
      ($\mathrm{R_{\odot}}$) & ($\mathrm{M_{\odot}}$) & ($\mathrm{K}$)
      & ($R_{*}$) & ($R_{*}$) & ($\mathrm{K}$) &
      ($\mathrm{\dot{M}_{\odot}yr^{-1}}$) & ($\mathrm{K}$) &
      ($\mathrm{km\, s^{-1}}$) & ($\mathrm{km\, s^{-1}}$)
      & $\cdots$ & ($R_{*}$)\\
      \hline
      $5.0$ & $2.8$ & $5200$ & $3.0$ & $3.8$ & $5500$ &
      $2\times10^{-6}$ & $9000$ & $530$ & $10$ & $1.8$ & $3.8$\\
      \hline
    \end{tabular}
  \end{center}
\end{table*}

%=========================================

The stellar luminosity and temperature of V1331~Cyg place the star on
the beginning of the radiative track for mass 2.8 M$_\odot$, in
between TTS and Herbig Ae stars, at the age of about 1.5 Myr.  Since
the star is oriented near pole-on to observer, it is hard to expect
any effect of rotational modulations. The irregular light variability
is probably caused by only the accretion processes. Thus, the period
of rotation remains unknown. Further, the small inclination angle of
V1331~Cyg implies that the LOS to the star (the
continuum radiation source) cannot intersect a disc
wind. Consequently, the wide and deep blueshifted wind absorption
feature seen in H$\alpha$ and H$\beta$ (Figs~\ref{balmer} and \ref{monitor}) are
not possible with this wind configuration. Most likely, such
absorption features are caused by a stellar wind that arises in the
polar directions (e.g.~\citealt{Edwards06}; \citealt*{Kwan07}; \citealt{Kurosawa11}).  For an
observer located in the polar direction, not
only the LOS to the stellar surface can easily pass through the
stellar wind, but also it can intersect with a full range of velocity
surfaces which provide a possibility to form the wide and deep
blueshifted absorption feature in the Balmer lines. For this reason, we
mainly focus on  modelling the observed line profiles using the bipolar
stellar wind and on probing its overall characteristics.  In the
following, we briefly describe our line profile models, and present
the comparison of our models with observed H$\alpha$ and H$\beta$ line
profiles.

\subsection{Model configuration}

%=========================================
\begin{figure}
\centering
\includegraphics[clip,width=0.49\textwidth]{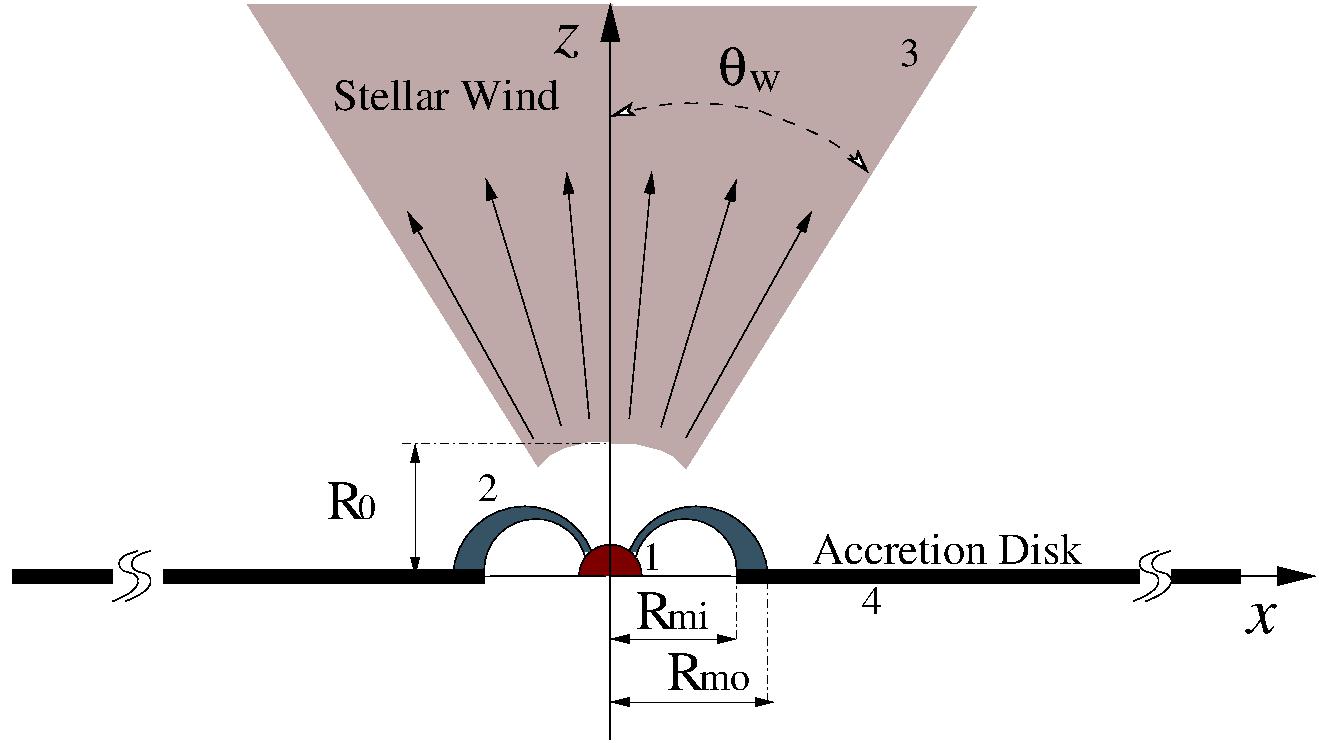}
 \caption{Basic model configuration of the stellar wind-magnetosphere
   hybrid model. The model is axisymmetric and it consists of four
   components: (1)~the continuum source (star) located at the origin
   of the Cartesian coordinates system $\left(x,z\right)$, (2)~the
  magnetospheric accretion flow, (3)~the (bipolar) stellar wind, and
  (4)~the optically thick but geometrically thin accretion disc.
  The wind is launched from a sphere with radius
  $R_{0}$, but is restricted within the cones with the half opening
  angle $\theta_{\mathrm{w}}$. The density distribution is symmetric
  around the $z$-axis. The figure is not to scale. }
\label{fig:model-config}
\end{figure}
%=========================================

To model emission line profiles of H$\alpha$ and H$\beta$, we use the
radiative transfer code {\sc torus} (e.g.~\citealt{Harries00, Kurosawa06, Kurosawa11}). 
In particular, the numerical method used in
the current work is essentially identical to that in \citet{Kurosawa11}. 
The model uses the adaptive mesh refinement
grid in Cartesian coordinate and assumes an axisymmetry around the
stellar rotation axis. The model includes 20 energy levels of hydrogen
atom, and the non-local thermodynamic equilibrium (non-LTE) level
populations are computed using the Sobolev approximation \citep{Sobolev57, Castor70}. 
For more comprehensive descriptions of the code, readers are referred to
~\citet{Kurosawa11}.

A basic schematic diagram of our model is shown in
Fig.~\ref{fig:model-config}. The model includes two flow components:
(1)~the dipolar magnetospheric accretion as described by \citet*{Hartmann94} 
and \citet*{Muzerolle01}, and
(2)~the stellar wind emerging from the polar regions. The radiation
from hotspots/rings formed on the stellar surface is also
included.  An optically thick and geometrically thin disc is placed on
the equatorial plane to imitate the absorption by the accretion disc.

The accretion stream through a dipolar magnetic field is described as
$r=R_{\mathrm{m}}\,\sin^{2}{\theta}$ (e.g.~\citealt*{Ghosh77}; \citealt{Hartmann94}) 
where $r$ and
$\theta$ are the polar coordinates; $R_{\mathrm{m}}$ is the
magnetopsheric radius at the equatorial plane. The accretion funnel
regions are defined by two stream lines corresponding
to the inner and outer magnetospheric radii, i.e.,
$R_{\mathrm{m}}=R_{\mathrm{mi}}$ and $R_{\mathrm{mo}}$. We adopt
the density and temperature structures along the stream lines as in
~\citet{Hartmann94}. The temperature scale is
normalized with a parameter $T_{\mathrm{m}}$ which sets the maximum
temperature in the stream.

The stellar wind is approximated as outflows in narrow cones with their
half-opening angle $\theta_{\mathrm{w}}$. Here, we assume the flow is
only in the radial direction, and its velocity is described by the
classical beta-velocity law (cf.~\citealt{Castor79}):
\begin{equation}
  v_{r}\left(r\right)=v_{0}+\left(v_{\infty}-v_{0}\right)\left(1-\frac{R_{0}}{r}\right)^{\beta}\,,
  \label{eq:beta-velocity-law}
\end{equation}
where $v_{\infty}$ and $v_{0}$ are the terminal velocity and the
velocity of the wind at the base ($r=R_{0}$). Assuming the mass-loss
rate by the wind is $\dot{M}_{\mathrm{w}}$ and using the mass-flux
conservation in the flows, the density $\rho_{\mathrm{w}}$
of the wind can be written as:
\begin{equation}
  \rho_{\mathrm{w}}\left(r\right)=\frac{\dot{M}_{\mathrm{w}}}{4\pi
    r^{2}v_{r}\left(r\right)\left(1-\cos\theta_{\mathrm{w}}\right)}\,.
  \label{eq:sw-density}
\end{equation}
Note that $\rho_{\mathrm{w}}$ becomes that of a spherical wind when
$\theta_{\mathrm{w}}=90^{\circ}$. The temperature of the stellar wind
($T_{\mathrm{w}}$) is assumed isothermal as in~\citet{Kurosawa11}.  
To avoid an overlapping of the stellar wind
with the accretion funnels, the base of the stellar wind ($R_{0}$) is
set approximately at the outer radius of the magnetosphere
($R_{\mathrm{mo}}$) (cf.~Fig.~\ref{fig:model-config}).

%=========================================
\begin{table}
  \begin{center}
    \caption{Model Parameters for Line Fits}
    \label{tab:model-par}
    \begin{tabular}{ccc}
      \hline
      Model ID & $\dot{M}_{\mathrm{w}}$ & $\theta_{\mathrm{w}}$\\
      $\cdots$ & ($\mathrm{\dot{M}_{\odot}yr^{-1}}$) & $\cdots$\\
      \hline
      A & $6.0\times10^{-8}$ & $50^{\circ}$\\
      B & $9.0\times10^{-8}$ & $50^{\circ}$\\
      C & $1.1\times10^{-7}$ & $50^{\circ}$\\
      D & $9.0\times10^{-8}$ & $40^{\circ}$\\
      E & $9.0\times10^{-8}$ & $60^{\circ}$\\
      \hline
    \end{tabular}
  \end{center}
\end{table}

%=========================================

\subsection{Balmer line models}
\label{sec:balmer-line-models}

The basic stellar parameters adopted for  modelling the observed
H$\alpha$ and H$\beta$ profiles of V1331~Cyg are:
$M_{*}=2.8\,\mathrm{M_{\odot}}$, $R_{*}=5\,\mathrm{R_{\odot}}$ and
$T_{\mathrm{eff}} = 5200$\,K, as found in
Section~\ref{sec:photospheric}.  Since we do not find a clear
periodic signature in the 20-d spectroscopic monitoring of
V1331~Cyg (Sect.~\ref{sec:wind-features}), we roughly estimate the
period by using $v\sin{i}<6\,\mathrm{km\,s^{-1}}$ and
$R_{*}=5.0\,\mathrm{R_{\odot}}$ (Sect.~\ref{sec:photospheric}). 
We assume a low inclination angle $i=10^{\circ}$, which sets lower limits
to the period of stellar rotation $P_*$ $\geq$ 7.4 d and the corotation radius 
$R_{\mathrm{cr}} \geq 4.5\,R_{*}$.
The inner and outer magnetospheric radii are set to
$R_{\mathrm{mi}}=3.0\,R_{*}$ and $R_{\mathrm{mo}}=3.8\,R_{*}$, which
are slightly smaller than the corotation radius.  Other important
model parameters adopted are summarized in
Table~\ref{tab:model-par-common}. Note that $P_{*}$ and $i$ used here
are rough estimates, and are only needed to find a reasonable size
of the magnetosphere. 

\begin{figure}
\centering
\includegraphics[clip,width=0.49\textwidth]{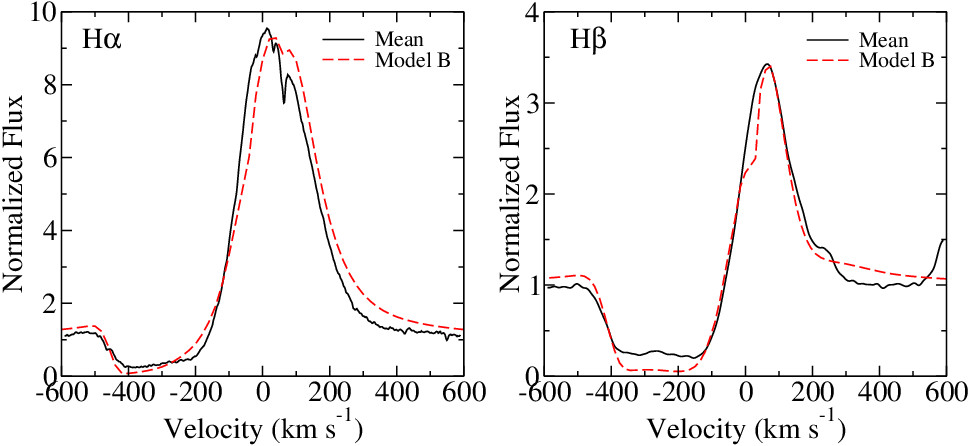}
 \caption{Comparisons of the mean H$\alpha$ and H$\beta$ profiles from
   the 2012 observations (solid; cf.~Table~\ref{Tab1}) with Model~B
   (dashed; Table~\ref{tab:model-par}). The matches between the model
   and the observations are excellent. The very wide and deep
   blueshifted absorption component (the P-Cyg line profile feature) is well
   produced with the bipolar outflow (the stellar wind) in our model.}
\label{fig:model-mean-prof}
\end{figure}

To model the line variability
behaviours seen in the observations (Figs.~\ref{balmer} and
\ref{monitor}),  we mainly concentrate on the effect of varying the
wind mass-loss rate ($\dot{M}_{\mathrm{w}}$) and the half-opening
angle of the bipolar stellar wind ($\theta_{\mathrm{w}}$) as in
Table~\ref{tab:model-par}.
To find a reasonable base model for the line variability, we first
fit the mean H$\alpha$ and H$\beta$ profiles from the 20-d
spectroscopic monitoring in 2012 (cf.~Table~\ref{Tab1}). The results
of the model fits are shown in Fig.~\ref{fig:model-mean-prof}. The
model uses $\dot{M}_{\mathrm{w}}=9.0\times
10^{-8}\,\mathrm{M_{\odot}\,yr^{-1}}$ and  $\theta_{\mathrm{w}}=50^{\circ}$
(Model~B in Table~\ref{tab:model-par}). Overall agreement of the model
and the observed mean profiles is very good. The model reproduces the
deep and wide absorption of the P-Cyg profiles very well.  
The line strengths and the widths are matched
well with the observations also. However, the wind absorption depth of
H$\beta$ tends to be slightly stronger in the model.
In this model, the emission component in both lines is mainly
originated in the wind. The contribution of the magnetosphere to the
line emission is much smaller, but it is important for producing a
slightly wider emission component than that from the wind emission alone.
This model (Model~B) is used as our base  for the line
variability  modelling which will be presented next. This model also sets
the common model parameter values given in
Table~\ref{tab:model-par-common}.

%=========================================
%=========================================

%=========================================
\begin{figure}
\centering
\includegraphics[clip,width=0.49\textwidth]{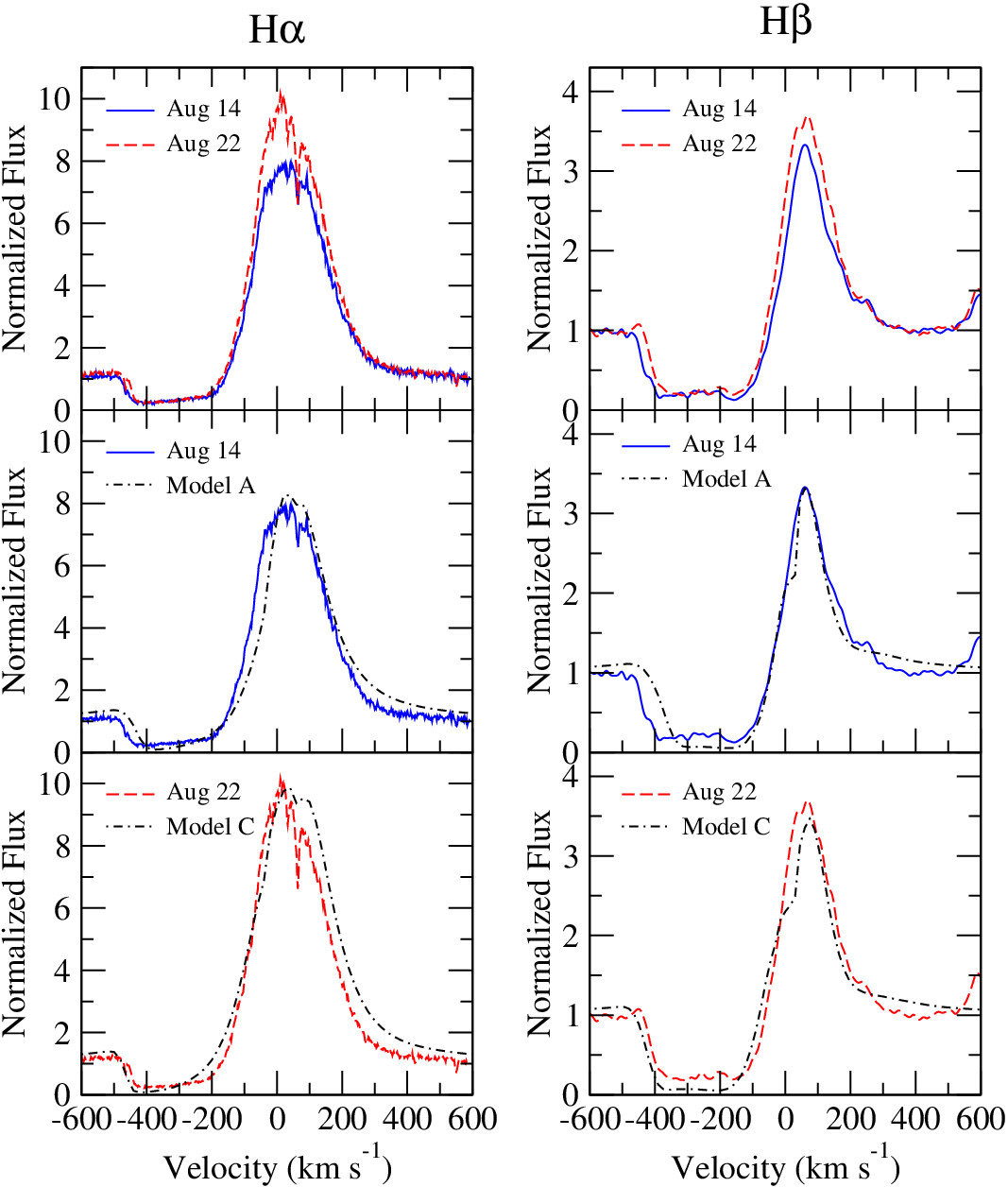}
 \caption{Top panels: the variability of H$\alpha$ and H$\beta$
   in about 8\,d
   time-scale, i.e. between Aug.~14 (solid) and Aug.~22 (dashed) in
   2012. Middle panels: the model line profiles (Model~A in
   Table~\ref{tab:model-par}) that fit the observation from 2012
   Aug.~14. Bottom panels: the model line profiles (Model~C in 
   Table~\ref{tab:model-par}) that fit the observation from 2012
   Aug.~22. The ranges of line variability seen in the observations at
   this 
   time-scale are well reproduced by changing the mass-loss rate in
   the bipolar stellar wind from 
   $6.0 \times 10^{-8}\,\mathrm{M_{\odot}\,yr^{-1}}$ (Model~A) to 
   $11 \times 10^{-8}\,\mathrm{M_{\odot}\,yr^{-1}}$ (Model~C).
}
\label{fig:model-var-week}
\end{figure}
%=========================================

Next, we examine the line variability that occurs in the time-scale of
about one week. For this purpose, we focus on the H$\alpha$ and H$\beta$
profiles observed at approximately 8~d apart, namely the data from
2012 Aug.~14 and Aug.~22 (the third and the last entries of Calar
Alto observations in Table~\ref{Tab1}).  The corresponding line
profiles are shown in the top panels of
Fig.~\ref{fig:model-var-week}. The figure shows that the peak
intensity of H$\alpha$ increases by a factor of 1.25, and that of
H$\beta$ increases by a factor of 1.1 during 8~d. The maximum
extent of wind absorption does not change significantly during this
time. In general, the line shapes do not change dramatically in this
time-scale (see also Fig.~\ref{monitor}), indicating the overall flow
structures of V1331~Cyg is stable in the time-scale of 
one week.

The relatively small variability seen in the observations is
well reproduced by our models (Models~A and B in the middle and lower
panels of Fig.~\ref{fig:model-var-week}) by changing the wind
mass-loss rate ($\dot{M}_{\mathrm{w}}$) slightly from our base model
(Model~B). The range of the mass-loss rates that fit the observed
variability in 8\,d period is $(6$--$11) \times 
10^{-8}\,\mathrm{M_{\odot}\,yr^{-1}}$. This relatively small change in the
mass-loss rate is perhaps caused by the response or adjustment of the
outflow to a change in the mass-accretion rate.  The fluctuation in
the mass-accretion rate by a factor of $\sim2$ naturally occurs and
is often observed in the MHD simulations of accretions on to cTTS
through a magnetosphere (e.g.~\citealt{Romanova02}).
Although we kept the mass-accretion rate in Models~A and C at a
constant value (Table~\ref{tab:model-par-common}), the effect of changing
the mass-accretion on H$\alpha$ and H$\beta$ line profiles is much
smaller than that of changing the wind mass-loss rates because the
wind emission dominates the observed line profiles, as we found
earlier.

\begin{figure}
\centering
\includegraphics[clip,width=0.49\textwidth]{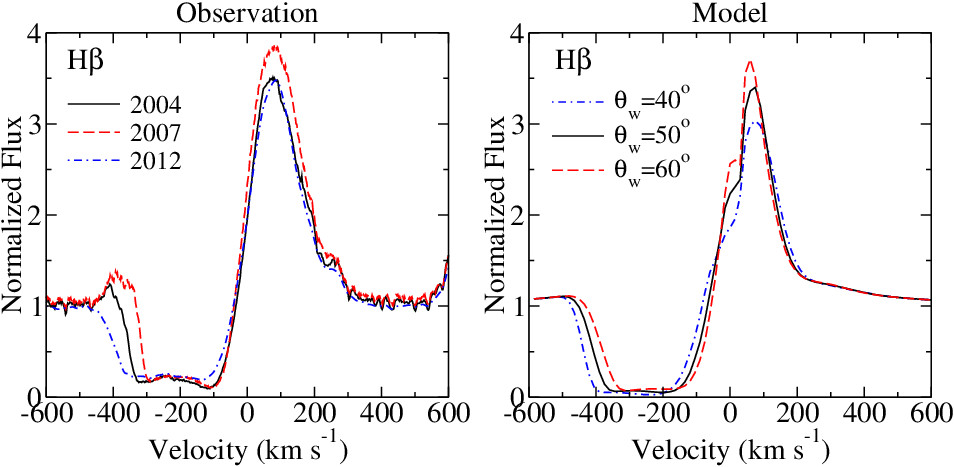}
 \caption{A comparison of the variability observed in H$\beta$ in the
   time-scale of several years (\emph{left panel}) with our radiative
   transfer models (\emph{right panel}). The variability in H$\beta$
   seen in this time-scale can be roughly reproduced by changing the
   half-opening angle ($\theta_{\mathrm{w}}$) of the bipolar stellar
   wind in our model between $40^{\circ}$ and $60^{\circ}$, i.e.,
   Models D and E in Table~\ref{tab:model-par}, respectively. The
   models show similar ranges of variability in the peak line flux
   and the extent (apparent maximum velocity) of the blueshifted wind
   absorption component as in the observations. N.B.\, the observed
   H$\beta$ shown in the left panel are same as those in
   Fig.~\ref{balmer}, but we omitted the data from 1998 for clarity.
}
\label{fig:model-var-yrs}
\end{figure}

As briefly mentioned earlier in Section~\ref{sec:wind-features}, the
variability of H$\beta$ in the time-scale of years (Fig.~\ref{balmer};
also in the left panel in Fig.~\ref{fig:model-var-yrs}) is also not so
very large. As shown in Fig.~\ref{fig:model-var-yrs}, the peak
intensity remains almost constant except for the data from 2007 which
is about 10 per cent higher than those of other years.  A more notable
variability is seen in the maximum velocity ($v_{\mathrm{max}}$) of
the blueshifted absorption component in the observed H$\beta$
profiles. From Fig.~\ref{fig:model-var-yrs}, we find
$v_{\mathrm{max}}$ changes approximately between -350 and -450
km\,s$^{-1}$.

This type of variability is well reproduced by adjusting
the half-opening angle ($\theta_{\mathrm{w}}$) of the bipolar wind between
$40^{\circ}$ and $60^{\circ}$ (Models~B, D and E
in Table~\ref{tab:model-par}), while keeping all other parameters
fixed.  The corresponding line profiles are shown and compared with
the observations in Fig.~\ref{fig:model-var-yrs}. The models show a
similar range of $v_{\mathrm{max}}$ values as in the
observations. Interestingly, in these models,
the terminal velocity of the wind ($v_{\infty}$ in
equation~\ref{eq:beta-velocity-law}) is fixed at the constant value of
$530\,\mathrm{km\,s^{-1}}$ which is higher than the $v_{\max}$ values.
Here, the change in the value of $v_{\mathrm{max}}$ in the models can be
understood by the change in the optical depth of the wind. Because the
wind mass-loss rate is fixed in these models, the density of the wind
increases as the half-opening angle ($\theta_{\mathrm{w}}$) of the
wind decreases (see equation~\ref{eq:sw-density}). Hence, the high optical
depth region in the wind, which causes the blueshifted absorption,
extends to a larger radius for a smaller
$\theta_{\mathrm{w}}$. This results in a larger value of
$v_{\mathrm{max}}$ or the \emph{apparent} terminal velocity. Note that
the value $v_{\mathrm{max}}$ can be smaller than $v_{\infty}$ when the
optical depth is significantly below 1 at outer radii where the wind
speed reaches $\sim v_{\infty}$.

Since very little is known about the formation process of the stellar wind in cTTS
itself, the physical cause of the change in the wind opening angle is
also unknown. Here, we speculate that the change in $\theta_{\mathrm{w}}$
may be caused by (1)~the change in the strengths of the open magnetic
field in the polar direction, and/or (2)~the change in the collimation
of an external wind such as the conical wind \citep{Romanova09} which can influence 
the flow geometry of the stellar wind.
 See Section~\ref{sec:discussion} for a further discussion on this
issue. 

In summary, our model with the bipolar stellar wind agrees well with
the general characteristics of the observed H$\alpha$ and H$\beta$
profiles from V1331~Cyg. Rather small variabilities seen in a week to
several-year time-scales can be reasonably reproduced by changing the
mass-loss rate and the opening angle of the stellar wind.

%=========================================

%%%%%%%%%%%%%%%%%%%%%%%%%%%%%%%%%%%%%%%%%%%%%%%%%%%%%%%%%%%%%%%%%%%
\section{Discussion}
\label{sec:discussion}

% marina
%\subsection{Discussion of the model, collimation of stellar wind}

Our studies show that a strong stellar wind is
required to explain the P Cyg features in Balmer lines of V1331~Cyg.
The comparisons of our models with the observations
show that this stellar wind is a necessary component of the flow.
It is not clear however, whether this wind can solely explain
the outflows observed at much larger distances from the star.
It may be possible that some other mechanisms of outflow also contribute to the
matter flux, such as the disc wind (e.g.~\citealt{Zanni07})
and conical wind from the disc--magnetosphere 
boundary (e.g.~\citealt{Romanova09}). The matter flux
in the stellar wind used in our models is about (3--5.5) per cent of the 
matter flux in the disc (see Tables~\ref{tab:model-par-common} and
\ref{tab:model-par}), which is within the range of those found in 
observations,  $\sim$(0.1--10) per cent (e.g.~\citealt{Hartigan95, Edwards06, Calvet98}); therefore, the matter flux used in our model could be  
sufficient to explain the large-scale outflows. However, the opening
angle of the stellar wind adopted in our model is relatively large
($\theta_{\rm w} = 40^\circ-60^\circ$), hence this wind should be somehow
collimated at larger distances, because the high-velocity jet component is usually 
well collimated. Alternatively, it is possible that
both stellar and inner disc winds contribute to the outflow.
Usually, the disc wind and conical wind have also a large opening
angle in the beginning of the flow, and hence they will not
restrict the wide-angle stellar wind, which is needed for
explaining the Balmer lines. However, these inner disc
winds may influence the collimation of the stellar wind.

In this study, we suggest that the variability seen in the Balmer
lines may be connected with changes of the opening angle $\theta_w$
due to inner disc winds (see Section~4.2).  For example, in case of
conical winds, the degree of collimation varies depending on the level
of magnetization $\sigma$ (ratio of the magnetic to matter pressure)
in the outflow. In case of low magnetization, $\sigma<0.01$ the
conical winds are only weekly collimated inside the simulation region
~\citep{Romanova09}, while in cases of higher
magnetizations, $\sigma\sim(0.1-0.3)$, the collimation is much stronger
~\citep*{Konigl11, Lii12}. Therefore, a small variation in the magnetic flux
threading conical winds may lead to a variation in the collimation of
stellar winds and consequently a variation of the shape of Balmer
lines.

Recent MHD models of stellar winds from cTTS
(e.g. \citealt{Matt05,Matt08}; \citealt{Cranmer09}) suggest that the
wind is `accretion-powered.' Their studies indicate that the mass-loss
and mass-accretion rates are coupled, i.e., the mass-loss rate would
increase if the accretion-rate increases.  On the other hand, the MHD
simulations by \citet{Romanova09} and \citet{Lii12} have shown that
the opening angle of the external wind (the conical wind) decreases
when the mass-accretion rate increases.

Combining the results from these studies, we expect the wind mass-loss
rate would become larger if the opening angle of the stellar wind
becomes smaller.  In our simple wind model
(equations~\ref{eq:beta-velocity-law} and \ref{eq:sw-density}), if the
mass-loss rate increases and the opening angle decreases, the density
of the wind would increase, assuming the velocity structure of the
wind does not change.  In general, if the mass-accretion rate
increases, the energy available to drive the wind would also increase;
hence, a stronger wind is expected to arise, i.e., with a higher
mass-loss rate and a higher terminal velocity.

In our line profile models (Section~\ref{sec:balmer-line-models}), we
found a change in the mass-loss rate can explain the variability on
small time-scales (days/weeks), but on larger time-scales (several
years), a change in the opening angle would play a more important
role.  As we have mentioned above, in reality a change in
mass-accretion rate and a change in opening angle might be coupled.
This may indicate that the changes in mass-accretion rates on shorter
time-scales (days/weeks) are much smaller (hence no/little change in
the opening angle) than those on longer time-scales (several
years). Since a change in the opening angle of the wind could be also
caused by a change in the strength of the stellar magnetic field, our
line profile analysis may suggest that the magnetic field strength is
relatively stable on small time-scales (days/weeks), but it changes
significantly in longer time-scales (several years).

%% discussion of emission line spectrum

The high mass-accretion rate, $\dot{M}_{\mathrm{a}}=2\times10^{-6}$
M$_\odot\,\mathrm{yr}^{-1}$, adopted in our model
(Section~\ref{sec:balmer-line-models}), places the star near the top end
of the full range of accretion rates observed in cTTS (e.g.~\citealt{Hartigan95, Edwards06}). 
%
% Based on the additional models we have computed (not shown here), we find the
% range of the mass-accretion rate, with which the model can still
% reasonably fit the mean H$\alpha$ and H$\beta$ profiles shown in Fig.~\ref{fig:model-mean-prof}, is
% $\dot{M}_{\mathrm{a}}=(1.5-2.5)\times10^{-6}\,\mathrm{M_{\odot}\,yr^{-1}}$.

  After finding the best fit model to the observed mean line profiles
  of H$\alpha$ and H$\beta$ (Fig.~\ref{fig:model-mean-prof}), we examined
  the sensitivity of the model to a change in mass-accretion
  rate. This was done to check the acceptable range of
  mass-accretion rates with which the model can reasonably fit the
  observed H$\alpha$ and H$\beta$ shown in
  Fig.~\ref{fig:model-mean-prof}. We find such range to be
  $\dot{M}_{\mathrm{a}}=(1.5$--$2.5)\times10^{-6}\,\mathrm{M_{\odot}\,yr^{-1}}$.

On the other hand, a slightly lower accretion rate follows from the observed
fluxes in H$\alpha$ and He\,{\sc i} 5876\,\AA\ emissions. Using the
empirical relationships between the line luminosities and mass
accretion rates found in~\citet{Rigliaco12} (see also
~\citealt*{Mohanty05, Herczeg08, Fang09}), we find
the mass-accretion rate of V1331~Cyg to be $\dot{M}_{\mathrm{a}} =
0.7(^{+1.5}_{-0.5}) \times10^{-6}$ M$_\odot\,\mathrm{yr}^{-1}$.  
Note that the effect of veiling is not included in this
estimate. In Section~\ref{subsec:veiling}, we found that the VF
between 4500 and 8500 \AA\ is rather uncertain,  
but the upper limit is VF$\leq$1.  
This means $\dot{M}_{\mathrm{a}}$
estimated from the line luminosities could be higher by a factor of up
to 2.  Considering the uncertainties, the mass-accretion rate used in
our model reasonably agrees with the value estimated from the line luminosity
measurements.  

The emission line spectrum of V1331 Cyg resembles that of the jet-driving Class~I
type young object V2492~Cyg \citep{Hillenbrand12}.
In both objects, besides the Balmer and other wind-sensitive lines, 
indicating intensive mass outflow, there are many permitted emission lines of low excitation,
neutral and singly ionized metals, which are relatively narrow ($\approx$50 km\,s$^{-1}$) 
and rested at stellar velocity. 
Interestingly, a weak Li\,{\sc i} 6707\,\AA\, emission was noticed in V2492 Cyg. 
It is also present in V1331 Cyg, being superposed with narrow photospheric 
absorption of the same transition. The appearance of Li\,{\sc i} 6707\,\AA\, in emission
at stellar velocity was noticed earlier in spectra of the FUor V1057 Cyg \citep{Herbig09}.
Apparently, in V1331 Cyg, there must be a volume of low-temperature, 
low-density gas which is not involved in the accretion/wind motions.
%There is no such a component in the standard model of the magnetospheric accretion.

%% discussion of shell lines

As shown in Section 3.6, the striking similarity in velocity profiles of the `shell' absorptions of metals and the forbidden emissions of [O\,{\sc i}] strongly suggests
the origin of the `shell' absorptions in the post-shocked gas in the jet, 
i.e. the jet is projected to the star. This might imply rather a small inclination
angle, provided the jet is straight and normal to the disc plane. However, in the 
[S\,{\sc ii}] image of V1331 Cyg vicinity, a wiggling jet was traced to as far
as 360 arcsec~\citep{Mundt98}, i.e. the jet is deviated of 
a straight line at large distances from the star. In our wind model we assume inclination
$i = 10^\circ$, but the resulted line profiles remain about the same even 
if the inclination is decreased by factor of 2.

The `shell' absorptions are not typical for cTTS but is common for the classical FUors 
V1057 Cyg and FU Ori ~\citep{Mundt84, Herbig03, Herbig09}. The two FUors inclined differently, 
so that a jet (if any) does not point to observer, but the wind is much stronger than in V1331 Cyg. Probably the `shell' lines in FUors are formed not in distinct expanding shells, 
but in the shocks within their powerful extended wind flows.
The case of V1331 Cyg is rare in a sense that the star is seen through its jet, 
so it may be considered as a {\it stellar analogue of blazar.}

%%%%%%%%%%%%%%%%%%%%%%%%%%%%%%%%%%%%%%%%%%%%%%%%%%%%%%%%%%%%%%%%%%%%%
\section{Conclusions}
\label{sec:conclusions}

From the analysis of the high-resolution spectra of the pre-FUor V1331
Cyg we conclude the following:
\begin{itemize}

\item the highly veiled photospheric spectrum belongs to G7-K0 IV star
  of mass 2.8 M$_\odot$ and radius 5 R$_\odot$. The intrinsic width of the photospheric
  lines is not resolved, $v\,\sin i$ $<$ 6 km\,s$^{-1}$, i.e. the star is seen pole-on.

\item the amount of veiling depends on line strength. The effect may be caused
  by abnormal structure of atmosphere heated by mass accretion.

\item the blue-shifted absorption of Fe\,{\sc ii}, Mg\,{\sc i}, K\,{\sc i} and some other metals form in a post-shocked gas within a jet.

\item the Balmer line profiles are reproduced by  model of bipolar stellar 
     wind with mass-loss rate (6--11)$\times$10$^{-8}$ M$_\odot\,\mathrm{yr}^{-1}$.
  
\item the Balmer line profile variabilities in several days to years time-scales 
  are reproduced by changes in mass-loss rate and opening angle of the stellar wind,
  which may be caused by small variations of magnetic flux threading the inner wind. 
  
\item in addition to the stellar wind, responsible for the observed
  P Cyg line profiles, the presence of conical wind and/or disc wind is
  suggested to explain the collimation at large distances.
\end{itemize}

In this work we considered only one specific case of a pre-FUor wind blowing
towards the observer, where the stellar wind component is dominant in formation of
the observed line profiles. 
It would be interesting to do similar study of wind(s) in a pre-FUor viewed at different inclination, e.g. LkHa 321, so that the disc wind (or conical wind) properties could 
also be investigated. 

\section*{Acknowledgements}
We thank the referee, Thomas Haworth, for valuable comments which
helped us improve the clarity of the manuscript.
Major parts of this work is based on the Keck spectra of V1331 Cyg
kindly provided to one of us (PP) by George Herbig in 2009.
We are grateful to Antonio Pedrosa for the spectrum taken in 1998.
 MF acknowledges financial support from grants AYA2011-30147-C03-01 of the 
Spanish Ministry of Economy and Competivity (MINECO), co-funded with EU 
FEDER funds, and 2011 FQM 7363 of the Consejer\'{\i}a de Econom\'{\i}a, 
Innovaci\'on, Ciencia y Empleo (Junta de Andaluc\'{\i}a, Spain).
Research of MMR was supported by NASA grant NNX11AF33G and 
NSF grant AST-1211318. RK thanks Suzan Edwards, Greg Herczeg and 
Stanislav Melnikov for valuable discussions.

%%%%%%%%%%%%%%%%%%%%%%%%%%%%%%%%%%%%%%%%%%%%%%%%%%%%%%%%%%%%%%%%%%%%%%

\label{lastpage}
\end{document}